\input{aipcheck.tex}
\documentclass[]{aipproc}
\layoutstyle{8d}
\SetInternalRegister\hbadness{8000}

\begin{document}

\title
      [Electron Temperature in Cas A]
      {The Electron Temperature and $^{44}$Ti Decay Rate in Cassiopeia A}

\author{J. Martin Laming}{
  address={Naval Research Laboratory, Code 7674L, Washington DC 20375, USA},
  email={jlaming@ssd5.nrl.navy.mil},
}

\copyrightyear  {2001}

\begin{abstract}
The effects of plasma elemental composition and ionization state on the effective
decay rate of $^{44}$Ti are investigated. We essentially follow the methods of
the first authors to treat this topic, Mochizuki et al., but use more realistic
plasma models, including radiative cooling, to compute the evolution of the charge
state distribution behind the reverse shock. For uniform density ejecta (i.e. no
clumps or bubbles) we find a negligible change to the decay rate of $^{44}$Ti. We
discuss the effects of non-uniform ejecta. We also briefly consider the effects on these
calculations of collisionless electron heating associated with weak secondary shocks
propagating throughout the Cas A shell as a result of foward or reverse shock
encounters with density inhomogeneities, recently suggested as an explanation for
the hard X-ray tail seen in BeppoSAX and RXTE/OSSE spectra.
\end{abstract}

\date{\today}

\maketitle

\section{Introduction}
Radioactive nuclei in the galaxy are mainly produced in supernova
explosions. Among the most important of these is $^{44}$Ti, which after
$^{56}$Ni and $^{56}$Co is the main energy source for the ejecta. Its
abundance is also sensitive to details of the explosion.
The observation of $\gamma$ rays from decay
products of $^{44}$Ti (i.e. $^{44}$Sc to which $^{44}$Ti decays
with a lifetime of $85.4\pm 0.9$ years \cite{ahmad00,gorres00}
and $^{44}$Ca, to which
$^{44}$Sc decays with a lifetime of a few hours)
in the Cassiopeia A supernova remnant by
the COMPTEL instrument on the Compton Gamma Ray Observatory \cite{iyudin94,iyudin97}
has sparked a re-examination of some of these ideas.
Cassiopeia A is the youngest known supernova remnant in the galaxy, with a likely
explosion date of 1680 A.D., and so is a good target to search for emission
from the decay products of $^{44}$Ti, since its
lifetime is a significant fraction of the age of the remnant.
The flux in the 1157 keV line of $^{44}$Ca of $4.8\pm 0.9\times 10^{-5}$
photons s$^{-1}$cm$^{-2}$ implies an initial mass of $^{44}$Ti of
$2.6\times 10^{-4}$ $M_{\odot}$\cite{iyudin94,iyudin97}.
The fluxes in the 67.9 and 78.4 keV
lines of $^{44}$Sc observed by OSSE and BeppoSAX are consistent with
lower values of the initial $^{44}$ Ti mass synthesized in the explosion
\cite{vink00}.
Neglecting the contribution of the continuum (a questionable assumption)
the observed flux is $2.9\pm 1.0\times 10^{-5}$ photons s$^{-1}$ cm$^{-2}$,
marginally consistent with the $^{44}$Ca flux.

The estimates of the $^{44}$Ti mass based on the $^{44}$Ca flux
are at the high end of the range suggested
by theory. Thielemann et al.
\cite{thielemann96} predict a $^{44}$Ti mass of $1.7\times 10^{-4} M_{\odot}$
for a 20$M_{\odot}$ progenitor, whereas other workers \cite{woosley95a,woosley95b,
timmes96} predict less than $10^{-4} M_{\odot}$. Since $^{44}$Ti is produced only in
the $\alpha$ rich freeze out (i.e. Si burning at low density, so that reactions
involving $\alpha$ particles have negligible rates \cite{arnett96}), Nagataki et al.
\cite{nagataki98} speculated that the yield of
$^{44}$Ti could be increased if the explosion were axisymmetric rather than spherically
symmetric, due to the existence for example of rotation or magnetic fields. The reason
for this is that in the reduced symmetry, $\alpha$ rich freeze out occurs at
higher entropy in the polar regions and hence produces more $^{44}$Ti relative to
$^{56}$Ni. An
aspherical explosion is also quite consistent with the morphology observed today in
the remnant \cite{fesen01}.

An equally ingenious solution was that of \cite{mochizuki99} who
suggested that $^{44}$Ti, if sufficiently highly ionized behind
the reverse shock (see below), would have its decay rate to
$^{44}$Sc reduced. This occurs because $^{44}$Ti decays mainly by
capture of a K-shell electron, and if ionized to the hydrogenic or
bare charge state, i.e. if the K shell electrons are removed, the
decay rate is reduced. In this paper we revisit the work of
\cite{mochizuki99} using some more recent ideas about the electron
heating in the Cas A ejecta to investigate the effect on the
$^{44}$Ti decay rate and the inferred mass of $^{44}$Ti.

\section{The Ionization State of the Cassiopeia A Ejecta}
\begin{figure}[htb]
    \includegraphics[height=.5\textheight]{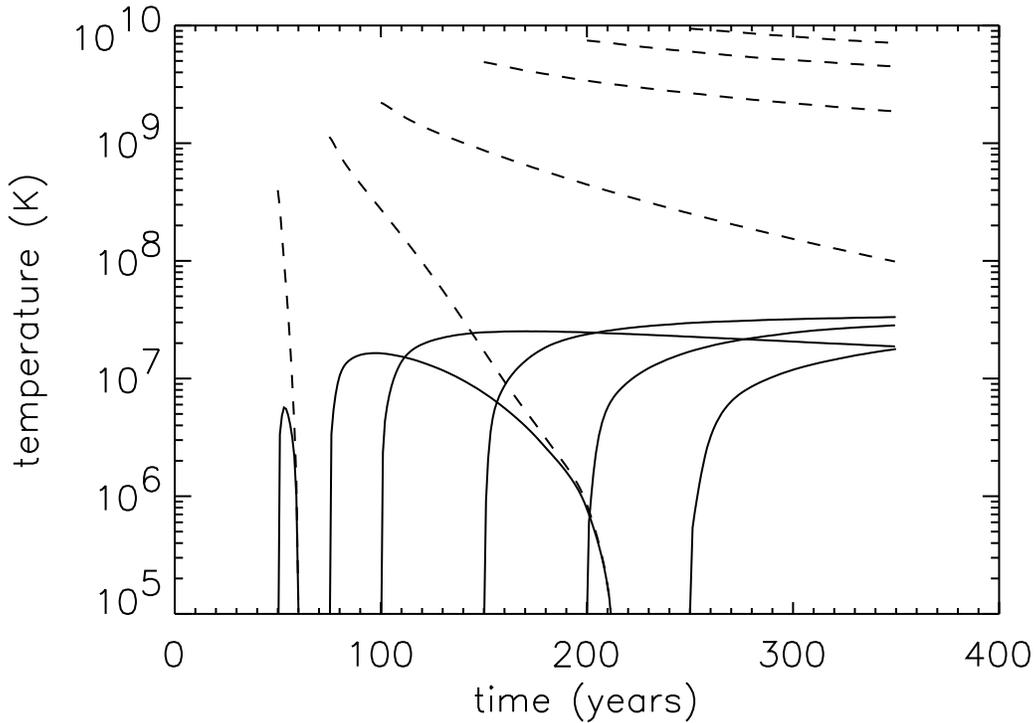}
  \caption{Electron (solid lines) and ion (dashed lines) temperatures in pure Fe ejecta
  passing through the reverse shock at 50, 75, 100, 150, 200, and 250 years after
  the initial explosion. The ejecta shocked at 50 years undergoes radiative
  cooling back to temperature below $10^5$ K within about 10 years, ejecta shocked at
  75 years takes 130 years to cool.}
  \label{fefig}
\end{figure}
\begin{figure}[htb]
    \includegraphics[height=.5\textheight]{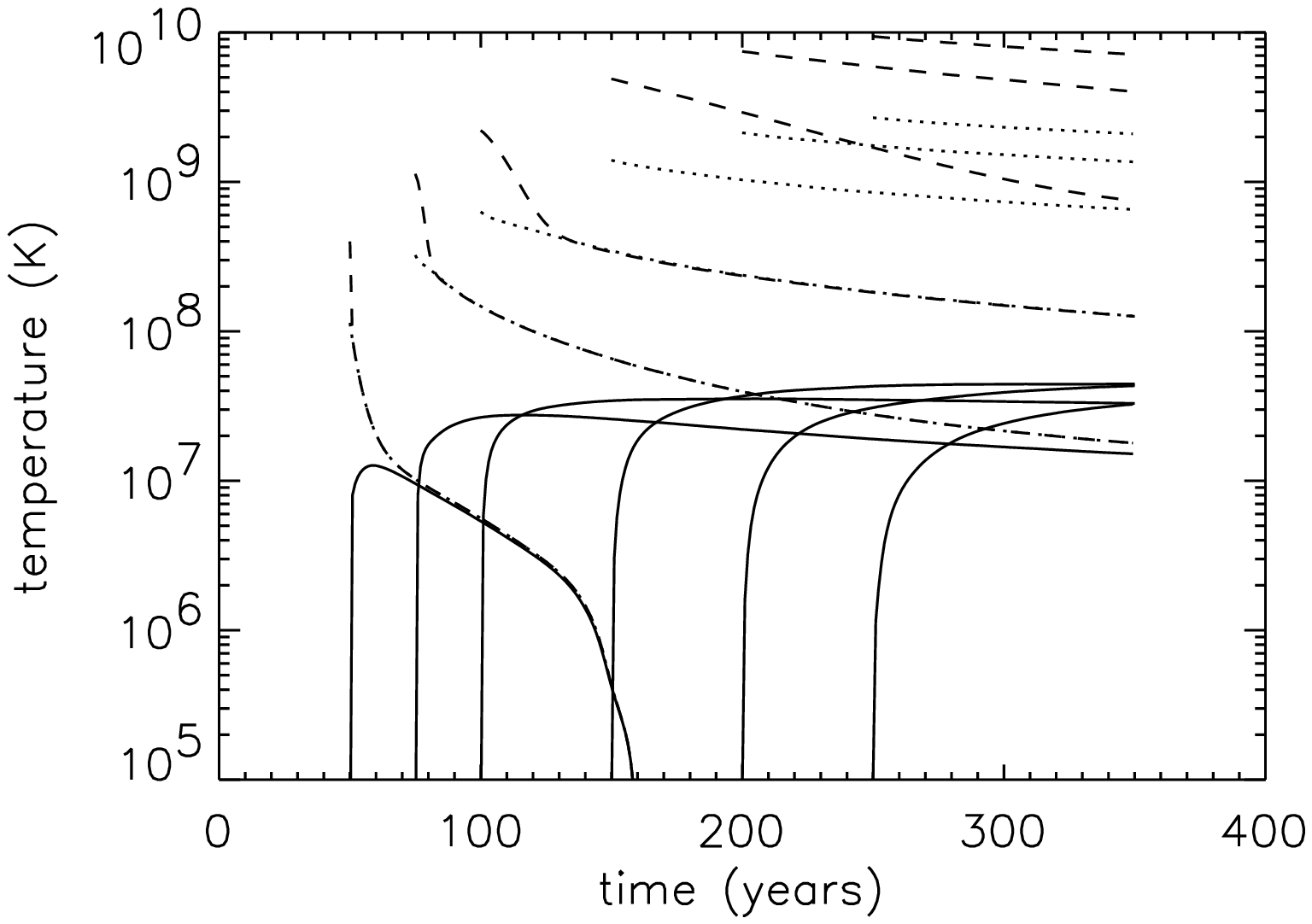}
  \caption{Electron (solid line), Fe ion (dashed line) and O ion (dotted line)
  temperatures in 90\% O, 10\% Fe (by mass) ejecta.}
  \label{ofefig}
\end{figure}
Following a supernova explosion, a spherical shock wave moves out through the
medium surrounding the progenitor (a presupernova stellar wind in the case of Cas A).
As this shock sweeps up more and more mass from the surrounding medium, it slows
down. This causes a reverse shock to develop, as freely expanding ejecta runs
into more slowly expanding shocked ambient plasma. This reverse shock moves
inwards in a Lagrangain coordinate system expanding with the plasma, and is responsible
for heating the stellar ejecta up to X-ray emitting temperatures. In investigations
of the element abundances produced by supernova explosions, the reverse shock is where
we focus our attention.

The ionization state of the Cas A ejecta following passage of the reverse shock
is computed using the formalism
described in \cite{laming02}. Assuming a total
ejecta mass of $4 M\odot$, an explosion energy of $2\times 10^{51}$ ergs and a
circumstellar medium density of 3 hydrogen atoms cm$^{-3}$, the reverse shock
velocity as a function of elapsed time since the explosion is taken from the
analytical expressions in \cite{mckee95}. Ionization and recombination rates
are taken from \cite{mazzotta98} (using subroutines kindly supplied by Dr P. Mazzotta).
Radiative cooling is taken from \cite{summers79}. Further cooling comes from the
adiabatic expansion of the ejecta. A well known problem is that the shocks in
supernova remnants are collisionless. Conservation of energy, momentum, and particle
number results in shock jump conditions that predict shocked particle temperatures
proportional to their masses, i.e. $T_{ion}/T_{electron} = m_{ion}/m_{electron}
\simeq 1836\times A_{ion}$ where $A_{ion}$ is the atomic mass.
These temperatures will equilibrate by Coulomb collisions, but the possibility
remains that collective effects may induce faster equilibration. For the time being
we neglect this possibility.

Figure \ref{fefig} shows the electron and ion temperatures
in pure Fe ejecta, for plasma encountering the reverse shock at times 50, 75, 100,
150, 200, and 250 years after explosion. Ejecta shocked 100 or more years after
explosion reaches electron temperatures of $2-3\times 10^7$ K, where appreciable
ionization of $^{44}$Ti to the H-like or bare charge states may occur. Ejecta
shocked after 75 years reaches an electron temperature of $2\times 10^7$ K briefly,
but cools by radiation and adiabatic expansion to temperatures an order of magnitude
lower at the present day. Ejecta shocked even earlier than this at 50 years never
reaches electron temperatures above $10^7$ K, and undergoes a thermal instability,
cooling catastrophically to temperatures below $10^5$ K within 15 years or so.
Mochizuki et al. \cite{mochizuki99} only included Fe charge states from Ne-like to
bare in their calculation (we use the full range of 27 charge states, starting
everything off in the Fe$^+$ state), and neglected to include radiation and so do not
find the thermal instability. Figure \ref{ofefig} shows similar plots but this
time of ejecta composed of 90\% O and 10\% Fe (by mass). Slightly lower electron
temperatures are found in the plasma shocked later in the evolution of the remnant,
but the reduced radiative cooling of this mixture gives higher electron temperatures
in the ejecta shocked at 50-75 years. The temperatures here are very similar
to those in pure O plotted in \cite{laming02}.

\begin{figure}[htb]
    \includegraphics[height=.5\textheight]{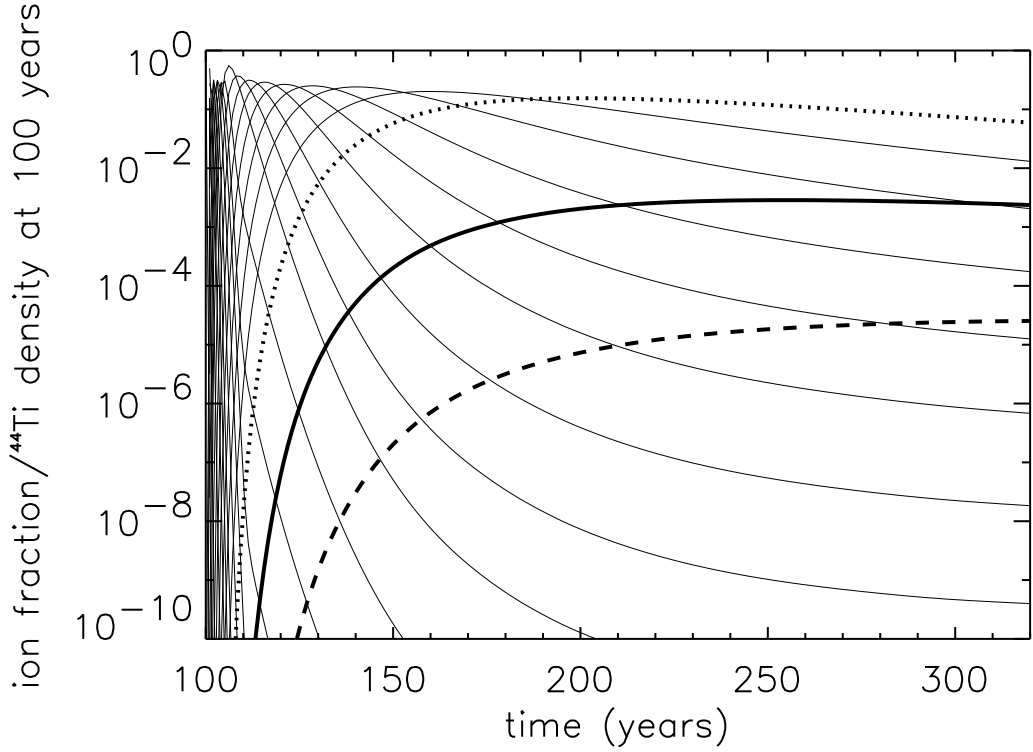}
  \caption{Ionization balance of $^{44}$Ti in pure Fe ejecta
  following reverse shock at 100 years. The
  He-like charge state is shown as the thick dotted line, the H-like as the thick
  solid line and the bare charge state as the thick dashed line. The narrow solid
  lines represent all the lower stages of ionization. At the present day (320 years)
  the He-like charge state dominates the ionization balance.}
  \label{44tifig}
\end{figure}

\section{$^{44}$Ti decay rate and radioactivity}
We now compute the ionization balance of trace amounts $^{44}$Ti assumed to be
expanding with the Fe ejecta, using the previously determined electron temperature
and density profiles. $^{44}$Ti decays by orbital electron capture, and the rate is
proportional to the electron probability density at the nucleus. Thus only $s$ states
contribute to the decay rate, and because the electron probability density at the nucleus
varies as $n^{-3}$, the $n=1$ shell is the dominant contribution. In the following
we assume that the $n=1$ shell is the {\em only} contribution. Thus all charge states
up to He-like will decay with the neutral decay rate, H-like will decay with half this
rate and the bare charge state will not decay at all. Figure \ref{44tifig} shows
the evolution of the Ti ionization balance following the reverse shock encounter
100 years after explosion. After a further 220 years, most of the Ti is in the He-like
charge state (the dotted line), but the smaller fraction that has accumulated in the
H-like (solid line) and bare (dashed line) charge states is sufficient to increase
the amount of Ti by a factor of 1.025 over that determined from the neutral decay rate
alone. The current emission in $^{44}$Sc and $^{44}$Ca is enhanced by a factor 1.009.
Results for various reverse shock encounter times are given in Table 1 for pure Fe
ejecta, and in Table 2 for mixed 90\% O, 10\% Fe (by mass) ejecta.
\begin{table}
\begin{tabular}{lrrr}
\hline
\tablehead{1}{r}{b}{time\\ (years)}
  & \tablehead{1}{r}{b}{mass\\ coordinate}
  & \tablehead{1}{r}{b}{$^{44}$Ti\\ density}
  & \tablehead{1}{r}{b}{$^{44}$Sc, $^{44}$Ca\\ emission}\\
\hline
50 & 0.72 & 1.0 & 1.0 \\
75 & 0.63 & 1.0 & 1.0 \\
100 & 0.55 & 1.025 & 1.009\\
150 & 0.43 & 1.001 & 1.0\\
200 & 0.34 & 1.0& 1.0\\
250 & 0.28 & 1.0& 1.0\\
\hline
\end{tabular}
\caption{Enhancements in $^{44}$Ti density and $^{44}$Sc and $^{44}$Ca emission
due to suppression of the $^{44}$Ti decay rate by plasma ionization. $^{44}$Ti
assumed embedded in pure Fe ejecta, for various reverse shock times after explosion.}
\end{table}
\begin{table}
\begin{tabular}{lrrr}
\hline
\tablehead{1}{r}{b}{time\\ (years)}
& \tablehead{1}{r}{b}{mass\\ coordinate}
  & \tablehead{1}{r}{b}{$^{44}$Ti\\ density}
  & \tablehead{1}{r}{b}{$^{44}$Sc, $^{44}$Ca\\ emission}\\
\hline 50 & 0.72 & 1.0 & 1.0 \\
75 & 0.63 & 1.11& 1.04 \\
100 & 0.55 & 1.09 & 0.995 \\
150 & 0.43 & 1.01& 1.002\\
200 & 0.34 & 1.0& 1.0\\
250 & 0.28 & 1.0& 1.0\\ \hline
\end{tabular}
\caption{Same as Table 1, but giving $^{44}$Ti enhancements etc in mixed 90\% O,
10\% Fe (by mass) ejecta.}
\end{table}
Mochizuki et al. \cite{mochizuki99} find a maximum enhancement of $^{44}$Ti radioactivity
of about 2.5 320 years after explosion for ejecta at mass coordinate $q=0.5-0.6$. We
find our maximum enhancement at the same mass coordinate, but only a factor of about
1.03 in pure Fe ejecta.
Mochizuki et al. take similar models to us, but in seeking to match a present
day blast wave velocity of $\sim 2000$ km s$^{-1}$ inferred in \cite{borkowski96},
adopted a circumstellar medium density much higher than ours.  We took a lower
circumstellar medium density to match the blast wave velocity of $\sim 5000$ km s$^{-1}$
found by \cite{vink98,koralesky98}. The effect of this change is that the
reverse shock in our model accelerates more slowly, generally producing lower
temperatures in the shocked ejecta, with correspondingly lower degrees of ionization
for the $^{44}$Ti. Mochizuki et al. also assumed
the Fe ejecta to expand in overdense (by a factor $\sim 10$) clumps, which reduces
the reverse shock velocity and temperature somewhat, but increases the electron density
and hence the ionization time, defined as the product of the electron density and
the elapsed time since shock passage, assuming an electron temperature constant with
time during this period.

\section{Further Electron Heating?}
The fundamental reason for not obtaining the same suppresion of the $^{44}$Ti decay
rate as Mochizuki et al.\cite{mochizuki99} is that the electron temperatures in our
probably more realistic model for Cas A do not reach high enough temperatures. Here we
will briefly discuss two possibilities so far neglected.

First, \cite{mochizuki99} assumed the Fe to exist in clumps a factor of 10 more dense
than the surrounding ejecta. In connection with Type Ia supernova, \cite{wang01}
argue that the Fe in such clumps is unlikely to have a radiogenic origin, and is probably
$^{54}$Fe. The reason is that the likely mechanisms to form clumps are hydrodynamic
instabilities following Ni-Co-Fe bubble formation as a result of the extra pressure
resulting from energy input mainly due to the Ni radioactivity in the first 10 days
following explosion.
Evidence for such bubbles is found in SN 1987A \cite{li93}, and some modeling has
already been performed \cite{borkowski00}. The $^{44}$Ti is likely
to reside in these bubbles, being formed in the same regions as $^{56}$Ni. If these
ejecta bubbles persist into the remnant phase, then the reverse shock will accelerate
into them due to their lower density. Higher temperatures in the shocked plasma will
result, with correspondingly higher degrees of ionization and possibly an effect
on the $^{44}$Ti decay rate.

For the time being it is not clear where in Cas A this Fe ejecta exists. We expect it
to be synthesized in the innermost regions of the progenitor, and so naively assuming
spherical symmetry we should expect the Fe to pass through the reverse shock relatively
late in the evolution of the remnant. However in the SE quadrant of the remnant, (the
region most extensively studied in Chandra data to date) the Fe rich knots are
remarkable for being {\em exterior} to those rich in Si and O
\cite{hughes00,hwang00,hwang01}.
These Fe rich knots also have high ionization times, implying high electron
density and/or early encounter with the reverse shock. The high density also
possibly suggests that this Fe is not associated with $^{44}$Ti. However the Fe rich
region gives mass ratios in the following ranges; Fe/O = 0.2 - 0.4, Si/O = 0.03 -0.06
\cite{hwang02}. Fe/O is consistent with or slightly higher than
with solar system values, while Si/O is lower. Earlier
X-ray observations, essentially spatially unresolved, of large regions of the remnant
find Si/O, S/O, Ar/O broadly consistent with solar system values, while Fe/O is
significantly lower \cite{vink96,favata97}.

The second possibility is that the electrons are heated following shock passage by
processes that are faster than the ion-electron Coulomb equilibration. Most discussion
of this in the literature has focused on plasma wave excited by instabilities at the
forward shock \cite[see e.g.][]{cargill88,bykov99,dieckmann00}.
At the reverse shock most electrons are bound to heavy ions, and so
cannot participate in collisionless heating \cite{hamilton84}.
However it is important to realize that
most of the ejecta emission comes from plasma that passed through the reverse shock
in the first 75 years or so after explosion. From Figures 1 and 2 it is apparent
that the temperature of these ejecta at the present day is rather lower than the
observed value of $\sim 4\times 10^7$ K, spectacularly so for the ejecta shocked at
around 50 years. This is due to the combined effects of radiation losses and
adiabatic expansion. Additionally, this plasma has
reached ion-electron temperature equilibrium, and so the inclusion of extra collisionless
processes at the reverse shock itself, however unlikely, would not produce any change to
this conclusion.

Recently it has been suggested that
this ejecta is reheated by weak secondary shocks produced as the forward or reverse
shocks run into density inhomogeneities producing reflected and transmitted shocks
\cite{laming01,laming02}. The precise mechanism is a modified two stream instability
arising as upstream ions are reflected from the shock front, move back through the
preshock plasma and excite lower-hybrid waves as they go. These waves are
electrostatic ion oscillations with wavevectors at angle nearly perpendicular to the
magnetic field direction. Thus the wave phase velocity perpendicular to the magnetic
field can resonate with the ions, and the wave phase velocity parallel to the field
can resonate with the electrons, allowing fast energy transfer between the two
\cite{bingham97,shapiro99}. Such a mechanism is particularly appealing in Cas A.
Arguing from the radio synchrotron luminosity, which is maximized when the
cosmic ray electrons and magnetic field have similar energy densities, leads to the
inference of magnetic field strengths of order 1 mG; significantly higher than usually
assumed for supernova remnants. The electrons are accelerated into a characteristic
distribution function along the magnetic field. Calculations of the bremsstrahlung
spectrum emitted by such an accelerated electron distribution produced a remarkably
good match to the BeppoSAX MECS and PDS spectra
\cite[see][for fuller description]{laming01}. In fact the computed continuum spectrum
for certain parameters (basically the Alfv\'en speed in the ejecta) can be consistent
with the seemingly extreme assumption made in \cite{vink00} of neglecting the
continuum under the $^{44}$Sc emission to derive a flux of $^{44}$Sc decay photons
consistent with the observation of the $^{44}$Ca flux. This accelerated electron
distribution equilibrates with the ambient plasma electrons by Coulomb collisions.
The extra heating thus produced would increase the ionization of the plasma above
that modeled in this paper, and could plausibly produce more $^{44}$Ti in H-like
or bare charge states, hence further suppressing its decay. However in order to avoid
heating the ejecta of Cas A to temperatures significantly higher than those observed,
such electron acceleration must have commenced relatively recently (i.e. within
50-100 years of the present day), if it is also responsible for the hard X-ray emission.

\section{Conclusions}
Cassiopeia A is an exciting physics laboratory for a wide variety of phenomena;
particle heating and acceleration, magnetic field amplification, hydrodynamic
instabilities and of course stellar nucleosynthesis. Initial ideas to study the element
abundances produced in a core-collapse supernova can often end up with
arguments based on observed element abundances to infer some other characteristic
of the supernova or its remnant. Examples of this are provided in this paper by the
plasma processes that may effect the decay rate of $^{44}$Ti. We find significantly
less change to the decay rate produced by the plasma ionization than previous work
\cite{mochizuki99}, but our model for the ejecta, while arguably more realistic, is
still probably a long way from reality itself.

X-ray astronomers have only just begun to get to
grips with the wealth of new data on Cassiopeia A and other supernova remnants, and
we can expect many exciting new results as this effort matures. This, coupled with
hard X-ray ray and $\gamma $-ray observations with the forthcoming INTEGRAL mission,
might finally allow to unambiguously infer exactly how much $^{44}$Ti was produced
in the explosion of Cassiopeia A.

\begin{theacknowledgments}
I acknowledge illuminating discussions and correspondence with Una Hwang and Jacco
Vink regarding the X-ray observations of Cas A.
This work was supported by basic research funds of the Office of Naval Research.
\end{theacknowledgments}

\bibliographystyle{aipproc}
\bibliography{casa_ti}
\end{document}